\documentclass[useAMS,usenatbib]{mn2e}

\usepackage{psfig}
\usepackage{amsmath}
\usepackage{amssymb}
\usepackage{upgreek}
\usepackage{longtable}
\usepackage[mathscr]{eucal}
\usepackage[dvips]{graphicx}

\title
[Kinematic signatures of cool collapse in NGC 6530]
{Kinematic signatures of cluster formation from cool collapse in the Lagoon Nebula cluster NGC 6530}
\author
[Wright et al.]
{Nicholas J. Wright$^1$\thanks{STFC Ernest Rutherford Fellow} and Richard J. Parker$^2$\thanks{Royal Society Dorothy Hodgkin Fellow}\\
\\
$^{1}$Astrophysics Group, Keele University, Keele, ST5 5BG, UK\\
$^{2}$Department of Physics \& Astronomy, The University of Sheffield, Hicks Building, Sheffield, S3 7RH, UK\\
}

\begin{document}
\maketitle

\begin{abstract}

We examine the mass-dependence of the velocity dispersion of stars in the young cluster NGC~6530 to better understand how it formed. Using a large sample of members we find that the proper motion velocity dispersion increases with stellar mass. While this trend is the opposite to that predicted if the cluster were developing energy equipartition, it is in agreement with recent N-body simulations that find such a trend develops because of the Spitzer instability. In these simulations the massive stars sink to the centre of the cluster and form a self-gravitating system with a higher velocity dispersion. If the cluster has formed by the cool collapse of an initially substructured distribution then this occurs within 1--2~Myr, in agreement with our observations of NGC~6530. We therefore conclude that NGC~6530 formed from much more extended initial conditions and has since collapsed to form the cluster we see now. This cluster formation model is inconsistent with the idea that all stars form in dense, compact clusters and provides the first dynamical evidence that star clusters can form by hierarchical mergers between subclusters.

\end{abstract}


\section{Introduction}

How star clusters form is one of the fundamental and outstanding questions in astrophysics. Young star clusters, particularly the most massive clusters, are typically dense and centrally concentrated, with a smooth radial distribution of stars \citep[e.g.,][]{carp00,pfal09}. This is very different from the distribution of dense gas in molecular clouds \citep{elme02,rath15} or very young stars in star-forming regions \citep{lars95,gute08}, both of which show a hierarchical and highly substructured spatial distribution. This implies that either the initial conditions for dense and centrally-concentrated star clusters are different to those observed in nearby, low-mass star forming regions \citep[i.e., such clusters form monolithically and in-situ from a highly-concentrated distribution of gas, e.g.,][]{bane14}, or that clusters assemble by hierarchical mergers between subclusters \citep{bonn03,mcmi07,alli09b,vazq17}, analogous to cold gravitational collapse models for galaxy formation \citep[e.g.,][]{vana82}. Distinguishing between these scenarios is important for understanding how clusters form, particularly young, massive clusters \citep{long14}.

The idea of star cluster formation by the merger of smaller sub-clusters fits well in the current model of star formation by turbulent fragmentation \citep{macl04}. Supersonic turbulence within molecular clouds leads to shocks that form filaments, dissipate energy \citep{ostr01} and lead to star formation, with the forming stars inheriting the subsonic motions of the gas from which they formed \citep{offn09}, as has been observed \citep{wals04,adam06,kirk07}. In their hydrodynamic simulations, \citet{bonn03} found that the forming stars were attracted by their mutual gravitational forces, falling towards each other and forming subclusters that then merged to form a single cluster. Similar simulations of subcluster mergers have been presented by \citet{mcmi07} and \citet{alli09b} to explain rapid dynamical mass segregation in young clusters, and by \citet{vazq17} to explain age spreads in star clusters, while \citet{long14} consider hierarchical mergers as one of the possible mechanisms for the formation of young massive clusters.

In this paper we present dynamical evidence that the young cluster NGC~6530 formed by hierarchical mergers from an initially substructured distribution. We find that the velocity dispersion of stars in NGC~6530 increases with stellar mass, a kinematic feature that recent N-body simulations have shown is developed when an initially substructured distribution of young stars undergoes cool collapse to form a single star cluster. In Section~2 we summarise the observational data used, in Section~3 we study how the velocity dispersion varies as a function of stellar mass, quantifying this in different ways. In Section~4 we compare our measurements with the predictions of N-body simulations and in Section~5 we discuss our results and their implications for star formation and the early evolution of young clusters.

\section{Observational data used}

\begin{figure}
\begin{center}
\includegraphics[height=175pt]{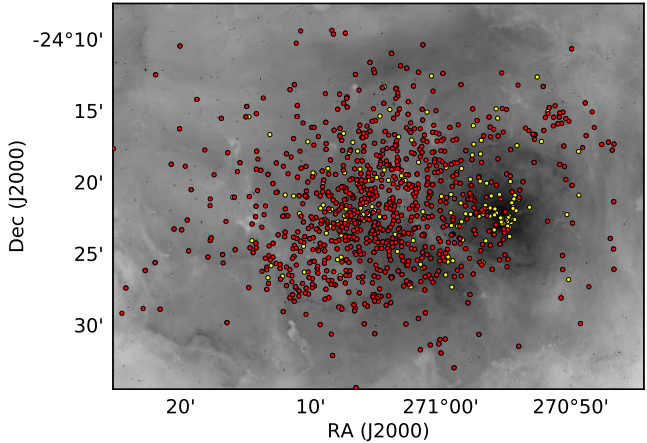}
\caption{Map of the spatial distribution of our kinematic sample of young stellar objects (red) and our spatially-unbiased `structural' sample of young stellar objects with $M > 0.7$~M$_\odot$, both projected onto an inverted H$\alpha$ image \citep{drew14}.}
\label{mini_map}
\end{center}
\end{figure}

The kinematic sample of young stars used for this study was compiled in \citet{wrig19} by combining spectroscopic information (surface gravity indicators, lithium equivalent widths and H$\alpha$ emission) from the {\it Gaia}-ESO Survey \citep[GES,][]{gilm12,rand13} with X-ray and infrared-based membership probabilities from \citet{broo13}. This results in a sample of $\sim$2000 high-probability members, of which $\sim$900 have reliable astrometry from {\it Gaia} \citep{prus16} data release 2 \citep[DR2,][]{brow18} that passed the data selection criteria outlined in \citet{aren18} and \citet{lind18b} for astrometric data. A parallax cut was then used to exclude the 135 stars more than 2$\sigma$ from the median parallax of the sample (0.724~mas or 1.33~kpc), and a further 70 stars were removed as proper motion (PM) outliers ($>$3$\sigma$ from the median PM in either dimension), providing a sample of 691 highly-probable young stars in NGC~6530 with PMs \citep[for more details on the compilation of this sample we refer the reader to][]{wrig19}.

In the work that follows we use the full {\it Gaia} covariance matrix when propagating uncertainties and consider all uncertainties to follow a normal distribution \citep{aren18}. We only consider the PM velocity dispersion for this study and not the radial velocity dispersion so as to limit the influence of the mass-dependent binary fraction and the impact unresolved binarity can have on radial velocity dispersions. Unresolved binaries are treated by {\it Gaia} as single stars, and so if their binary motion is significant this will lead to corrupt astrometry that might affect the measured proper motions. However, this will also mean that the measured astrometry would not be fit well by the 5 parameter astrometric model used by {\it Gaia} \citep[e.g.,][]{aren18} and the source would therefore be rejected by our astrometric quality cuts.

In addition to this kinematic sample of young stars we also compiled a more extensive sample of high-probability young stars in NGC~6530 that will be used for a structural study of the spatial distribution of young stars similar to that performed by \citet{wrig14b}. This sample was based on the sample of $\sim$2000 high-probability members already compiled, but without the requirement that they had reliable astrometry (though if a source did have a reliable parallax this was still used to validate membership). This provided a sample of $\sim$1900 highly-probable young stars in NGC~6530.

For both samples, stellar masses for all stars were taken from \citet{wrig19}, calculated from comparison of the available photometry to \citet{mari17} stellar isochrones. The stars in both samples have stellar masses that vary from $\sim$100~M$_\odot$ for the O4V star HD~164794 down to $\sim$0.1~M$_\odot$ for the faintest stars.

For the spatial sample, we note that X-ray observations can have a spatially-varying sensitivity that can affect the detection of low-mass stars \citep[see e.g.,][]{wrig15b} and therefore bias studies of the spatial distribution of stars in a star forming region or cluster. \citet{wrig19} estimated that the mass function of highly-probable members of NGC~6530 exhibited a turnover at around 0.7~M$_\odot$ and we use this as the estimated completeness limit of our sample, removing all stars from our structural sample with masses below this and leaving a sample of $\sim$800 stars. The spatial distribution of both samples are shown in Figure~\ref{mini_map}.

\section{Results and analysis}
\label{s-velocitydispersion}

In this Section we study how the velocity dispersion in NGC~6530 varies as a function of stellar mass. We explore this by measuring the velocity dispersion binned by stellar mass (Section~\ref{s-veldisp_by_mass}), by performing a global velocity dispersion fit with a mass-dependent term (Section~\ref{s-equipartition_fit}), and by calculating the ratio of velocity dispersions of high- and low-mass stars (Section~\ref{s-veldisp_ratio}). All of these methods show that the high-mass stars are typically moving more rapidly than the low-mass stars. While filtering the data on the {\it Gaia} DR2 `astrometric excess noise' quantity is not recommended \citep{lind18}, we repeated all the analysis that follows with only those stars with an astrometric excess noise of zero and found that our results did not change.

\subsection{Velocity dispersion as a function of stellar mass}
\label{s-veldisp_by_mass}

To determine how the velocity dispersion varies as a function of stellar mass we divided our sample into subsets based on mass and then calculated the PM velocity dispersion for each. Due to the form of the initial mass function the vast majority of our sample have very similar masses, within approximately 0.7--2.0~M$_\odot$. This means that dividing our full sample into equally-sized sub-samples would not allow us to accurately probe a wide range of stellar masses. Instead we divided our sample by stellar mass into four differently-sized samples that would allow us to probe how the velocity dispersion varies as a function of stellar mass. We note that binning the sample in this way (or any way for that matter) is not ideal, but is necessary for this approach.

We measure the velocity dispersion of stars in each group using the inter-quartile range (IQR), an outlier-resistant method for measuring the width of a distribution. The IQR of a Gaussian distribution of velocities is related to the velocity dispersion by $\sigma = 0.741 \times \mathrm{IQR}$. We account for the contribution of non-uniform PM uncertainties on the velocity dispersion using the outlier-resistant method of \citet{ivez14}, and determine the uncertainties on the resulting velocity dispersions by bootstrapping. The velocity dispersion was calculated for each PM dimension and then combined to produce a 2D velocity dispersion, $\sigma_{2D}$.

\begin{figure}
\begin{center}
\includegraphics[height=230pt, angle=270]{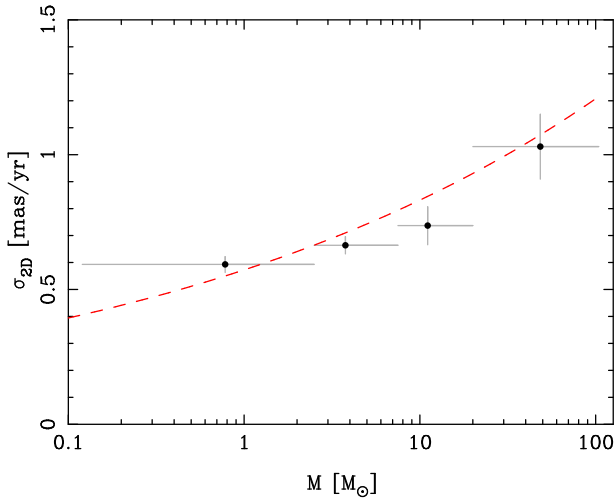}
\caption{2D PM velocity dispersion, $\sigma_{2D}$, as a function of stellar mass. Vertical error bars show the 1$\sigma$ uncertainty in the velocity dispersions and the horizontal error bars show the full distribution of stellar masses in each bin (bins span masses of 0.1--2.5, 2.5--7.5, 7.5--20, and 20--100 M$_\odot$ and contain 595, 79, 11 and 6 stars, respectively). The red dashed line shows the best-fitting value of $\eta = -0.081$ obtained by fitting the entire sample with a mass dependent velocity distribution where $\sigma (m) = \sigma_0 \, m^{-\eta}$.}
\label{mass_dependence}
\end{center}
\end{figure}

Figure~\ref{mass_dependence} shows the 2D velocity dispersion as a function of stellar mass. There is a clear trend of increasing velocity dispersion towards higher stellar masses, which is particularly significant for the most massive stars. Variations in the position and width of the bins (and therefore of the number stars in each bin), did not result in significant changes to these results. Decreasing the number of stars in each bin had the effect of increasing the uncertainty on the measured velocity dispersion. Increasing the number of stars decreased the uncertainty, but had the effect of blurring the trend between mass and velocity, particularly at high mass (for example if the number of stars in the highest-mass bin is doubled then the mass range changes to 10--100~M$_\odot$ and the velocity dispersion drops by $\sim$10\%, approximately mid-way between the previously two highest bins).

\subsection{Global mass-dependent velocity dispersion fit}
\label{s-equipartition_fit}

An alternative approach to measure the mass-dependence of the velocity dispersion (and to avoid any complications or biases arising from binning the data) is to fit a global velocity dispersion model to the PM distributions with a mass dependent term, such that

\begin{equation}
\sigma \, (m) = \sigma_0 \, m^{-\eta}
\end{equation}

\noindent where $\sigma$ is the velocity dispersion for stars of mass $m$ in a given dimension, $\sigma_0$ is the velocity dispersion for 1~M$_\odot$ stars in that dimension, and $\eta$ is the mass-dependence of the velocity dispersion. This formulation follows \citet{tren13}, who fitted this function to determine the level of energy equipartition in the system (where $\eta = 0$ is no mass dependence on the velocity dispersion and $\eta = 0.5$ is full energy equipartition). While energy equipartition actually involves the most massive stars moving {\em slower} than the less-massive stars, this formulation is ideal for our needs.

We fit this model using Bayesian inference and use the Markov-Chain Monte Carlo (MCMC) ensemble sampler {\em emcee} \citep{fore13} to sample the posterior distribution, using the method of \citet{wrig19}. We used 1000 walkers and 2000 iterations, with the first 1000 iterations used to explore the parameter space (the `burn-in') and the second 1000 iterations used to sample the posterior distribution. We used the median value of the resulting posterior distribution as the best fit, and the 16$^{th}$ and 84$^{th}$ percentiles for the 1$\sigma$ uncertainties. The longest autocorrelation length of the walker chains was found to be $\sim$120 iterations, resulting in $\sim$8 independent samples per walker.

The best fit was $\eta = -0.081 \pm 0.029$, which implies with 2.8 $\sigma$ confidence that the more massive stars in NGC~6530 are moving faster than the low-mass stars. This relationship between stellar mass and velocity dispersion is shown in Figure~\ref{mass_dependence}, in broad agreement with the binned data shown in that figure and confirming that the trend observed in Section~\ref{s-veldisp_by_mass} is independent of the choice of bins. This is in contrast to \citet{wrig16} who used this approach studying the Cygnus~OB2 association and found no dependence of the velocity dispersion on stellar mass in their sample.

\subsection{Velocity dispersion ratio for massive stars}
\label{s-veldisp_ratio}

To quantify the effect seen in Figure~\ref{mass_dependence} and to facilitate reliable comparisons with simulations we divide the velocity dispersion of the $N$ most massive stars, $\sigma_\mathrm{massive}$, with the velocity dispersion of all stars in the sample, $\sigma_\mathrm{all}$, to define a `velocity dispersion ratio', $\sigma_\mathrm{VDR}$ :

\begin{equation}
\sigma_\mathrm{VDR} = \frac{\sigma_\mathrm{massive}}{\sigma_\mathrm{all}}
\label{eq:vdr}
\end{equation}

Uncertainties for this quantity can be calculated, as in Section~\ref{s-veldisp_by_mass} by bootstrapping. A value of this ratio significantly above (below) 1.0 implies that the massive stars in the region are moving significantly faster (slower) than the low-mass stars. It can therefore be used to either quantify trends such as those observed here, or in more dynamically evolved groups, such as globular clusters, the development of energy equipartition.

We measure the velocity dispersion ratio for our sample for different values of $N$ (in steps of 5), exploring the effects of varying this quantity. We find that for $N \leq 30$ the velocity dispersion ratio is always greater than 1.0 and increases as $N$ decreases. The significance of the measurement broadly decreases as $N$ decreases, implying that the ratio increases faster than the uncertainty on the ratio increases. The highest value of the ratio was found for $N = 5$, where $\sigma_\mathrm{VDR} = 2.02 \pm 0.31$, implying that the most massive stars are moving, on average, {\em twice} as fast as the average star with a confidence of $\sim$3$\sigma$. The velocity dispersion ratio decreases for larger $N$, though the uncertainty initially decreases faster, such that for $N = 15$ the velocity dispersion ratio is measured as $\sigma_\mathrm{VDR} = 1.56 \pm 0.16$, a significance of 3.5$\sigma$. For larger $N$ the velocity dispersion ratio decreases, while the uncertainty remains constant, and so the significance of the measurement also drops.

\begin{figure*}
\begin{center}
\includegraphics[height=500pt, angle=270]{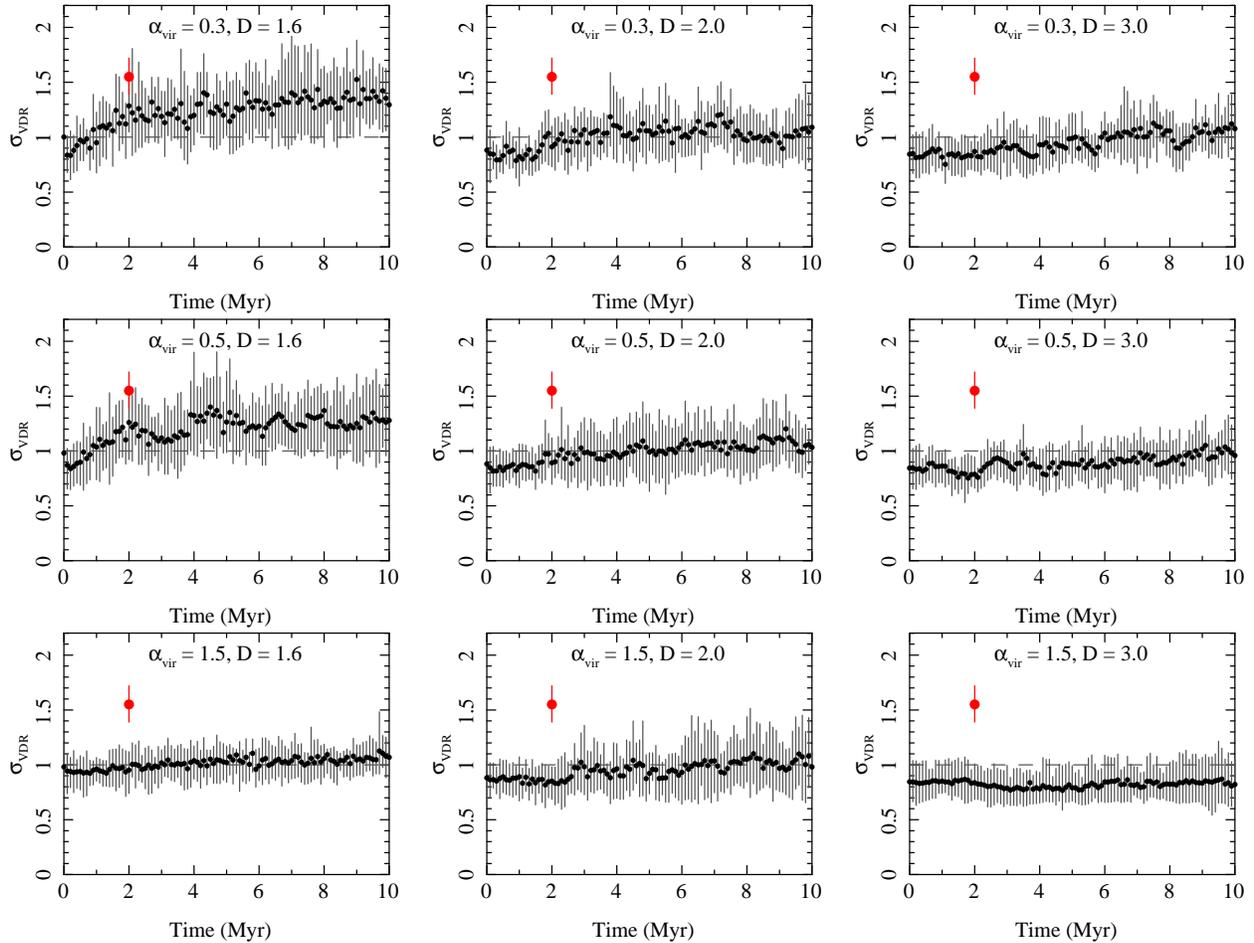}
\caption{Evolution of the velocity dispersion ratio, $\sigma_\mathrm{VDR}$ with $N = 15$, for stars within 2 half-mass radii of the centre of each region. Each panel shows the median $\sigma_\mathrm{VDR}$ value from 20 simulations with identical initial conditions (black dots) at 0.1~Myr time-steps, with the error bars showing the 16th and 84th percentile values of the $\sigma_\mathrm{VDR}$ values at that time-step. The three rows show initial conditions with a `cool' initial viral state ($\alpha_\mathrm{vir} =0.3$, top), a `virial' ratio ($\alpha_\mathrm{vir} = 0.5$, middle), and a `warm' initial virial state ($\alpha_\mathrm{vir} = 1.5$, bottom). The three columns show initial conditions with a high level of initial substructure ($D = 1.6$, left), a medium level of initial substructure ($D = 2.0$, centre), and no initial substructure ($D = 3.0$, right). The red dot shows the measured value of $\sigma_\mathrm{VDR} = 1.55 \pm 0.16$ for $N = 15$ for our NGC~6530 sample at an assumed age of 2~Myr \citep{bell13}.}
\label{ee_simulations}
\end{center}
\end{figure*}

\subsection{Consideration of possible biases and uncertainties}

It is worth considering any possible biases or uncertainties that could affect our results. The most important thing to consider is whether over- or under-estimated PM uncertainties could have significantly affected these results. Under-estimated uncertainties will cause the inferred velocity dispersion to be over-estimated (and vice-versa). We have calculated the extent to which the uncertainties need to be over- or under-estimated for this to explain the observed trend and find that this is incredibly unlikely. For the lowest-mass stars (those in the 0.1--2.5~M$_\odot$ bin), their uncertainties would need to be over-estimated by $\sim$0.6~mas~yr$^{-1}$ to bring their velocity dispersion to the level of the high-mass stars, which is impossible since the rms PM uncertainty for the low-mass stars is only 0.26~mas~yr$^{-1}$. For the highest-mass stars (20--100~M$_\odot$) their uncertainties would need to be under-estimated by a similar amount to bring their velocity dispersion down to the level of the lowest-mass stars, or even under-estimated by $\sim$0.5~mas~yr$^{-1}$ to bring it down to the level of the next-highest bin (7.5--20~M$_\odot$), which is a factor 15 larger than the rms PM uncertainty of the high-mass stars (0.03~mas~yr$^{-1}$).

Another possibility is that our sample is incomplete in some way that biases the velocity distribution of the sample, e.g., due to obscuration of low-mass stars with large velocities in the H{\sc ii} region that the stars are projected against, though this is hard to quantify and would also have to be a large effect to explain these observations.

\section{Comparison with N-body simulations}
\label{s-simulations}

In this section we compare our findings with the predictions of N-body simulations to explore how the observed kinematic trend could have arisen and how it might constrain the initial conditions or past evolution of the region.

We use the N-body simulations presented by \citet{park14} and subsequently studied by \citet{park16}. These simulations are discussed in detail in the first of those papers, but we summarise their properties here. The regions simulated have 1500 members drawn from a \citet{masc13} initial mass function with $\alpha = 2.3$ and $\beta = 1.4$, which results in a total mass of $\sim$500~M$_\odot$. The simulations are run from nine sets of initial conditions representing all combinations of three levels of initial physical substructure (with fractal dimensions $D = 1.6$, 2.0 and 3.0, in order of decreasing substructure) and three levels of initial virial ratio ($\alpha_\mathrm{vir} = T / | \Omega | = 0.3$, 0.5, and 1.5, referred to as `cool', `virial', and 'hot', respectively). For each set of initial conditions an ensemble of 20 simulations are run with different random number seeds used to initialise the positions, masses and velocities of the stars. The simulations are run for 10~Myr using the {\sc starlab} package \citep{port99} with the positions and velocities of stars outputted at 0.1~Myr intervals for analysis.

\subsection{Velocity dispersion ratio}
\label{s-vdr_simulations}

\citet{park16} analysed the kinematics of stars in the simulations of \citet{park14}. One of their notable findings was that in regions undergoing cool collapse (i.e., those starting from substructured initial conditions, $D = 1.6 - 2.0$, and cool or virial kinematics, $\alpha_\mathrm{vir} = 0.3 - 0.5$), the most massive stars had higher velocity dispersions than the low-mass stars after $\sim$1--2\,Myr of dynamical evolution. This was the case for both the 10 most-massive stars, which they argued may have kinematically `decoupled' from the other stars, and for all intermediate mass stars (1--5~M$_\odot$).

\citet{park16} found that this inflated velocity dispersion did not seem to be due to the high-mass stars being in close binary systems (which would increase their velocity dispersion in an N-body simulation if they were considered individually), as the majority of massive stars in the simulation did not end up in close binary systems. We remind the reader that while the high-mass stars in NGC~6530 may be in close binaries, our use of PMs instead of radial velocities means that the measured velocities are not instantaneous measures but are integrated over time, erasing the majority of binary motions.

To compare our results with these simulations we calculate the velocity dispersion ratio outlined in Section~\ref{s-veldisp_ratio} every 0.1\,Myr in each of the simulations, using $N = 15$ as a balance that provides a reasonably-sized sample to calculate the velocity dispersion of the most massive stars, without going too far down the mass spectrum that the signal is diluted. We limit the stars analysed at each time-step to those within 2 half-mass radii of the centre of each region so that the N-body simulations (which follow the evolution of all stars) mimic the observations (that are spatially limited to the main concentration of stars).

Figure~\ref{ee_simulations} shows how the velocity dispersion ratio, $\sigma_\mathrm{VDR}$ (Eq.~\ref{eq:vdr}), varies with time for various initial conditions. The ratio starts around 0.8--1.0, and with the exception of the warm initial conditions without substructure ($\alpha_\mathrm{vir} = 1.5$, $D = 3.0$), the ratio increases over time. The rate of increase of the velocity dispersion ratio is higher for simulations with more initial substructure and for simulations that start with `cool' or `virial' velocities. In the simulations with highly-substructured initial conditions ($D = 1.6$) that undergo a cool collapse ($\alpha_\mathrm{vir} = 0.3$ and 0.5), the velocity dispersion ratio increases quickly in the first few Myrs before flattening off to a level of 1.3--1.5. A slightly similar effect is seen in the partially-substructured simulation that undergoes cool collapse ($\alpha_\mathrm{vir} = 0.3$, $D = 2.0$), though the initial increase in the velocity dispersion ratio is less pronounced and is around 1.0--1.2 after 10~Myr. For the remaining simulations (with the exception of those with warm initial conditions without substructure) the velocity dispersion ratio increases much more slowly over the duration of the simulations, reaching approximately 1.0--1.1 after 10~Myr. In summary, and in agreement with \citet{park16}, the velocity dispersion ratio increases quickly for simulations undergoing cool collapse, while the ratio increases much more slowly for regions that dynamically evolve slower.

Comparing our measurement of the velocity dispersion ratio in NGC~6530 at an age of 2~Myr\footnote{Note that some more recent studies have derived a younger age for NGC~6530 of 1--2~Myr \citep[e.g.,][]{pris19}, which only accentuates the results of our comparison with simulations.} \citep{bell13} with the simulations shown in Figure~\ref{ee_simulations} shows that our observations are only consistent with simulations that undergo a cool collapse ($\alpha_\mathrm{vir} = 0.3$ or 0.5) from an initially highly-substructured distribution ($D = 1.6$). For all other initial conditions the velocity dispersion ratio does not reach such levels by an age of 2~Myr, and only for a few sets of initial conditions (e.g., the partially-substructured distributions that undergo cool collapse, where $D = 2.0$ and $\alpha_\mathrm{vir} = 0.3$) does it reach such a level within 10~Myr.

\begin{figure*}
\begin{center}
\includegraphics[height=500pt, angle=270]{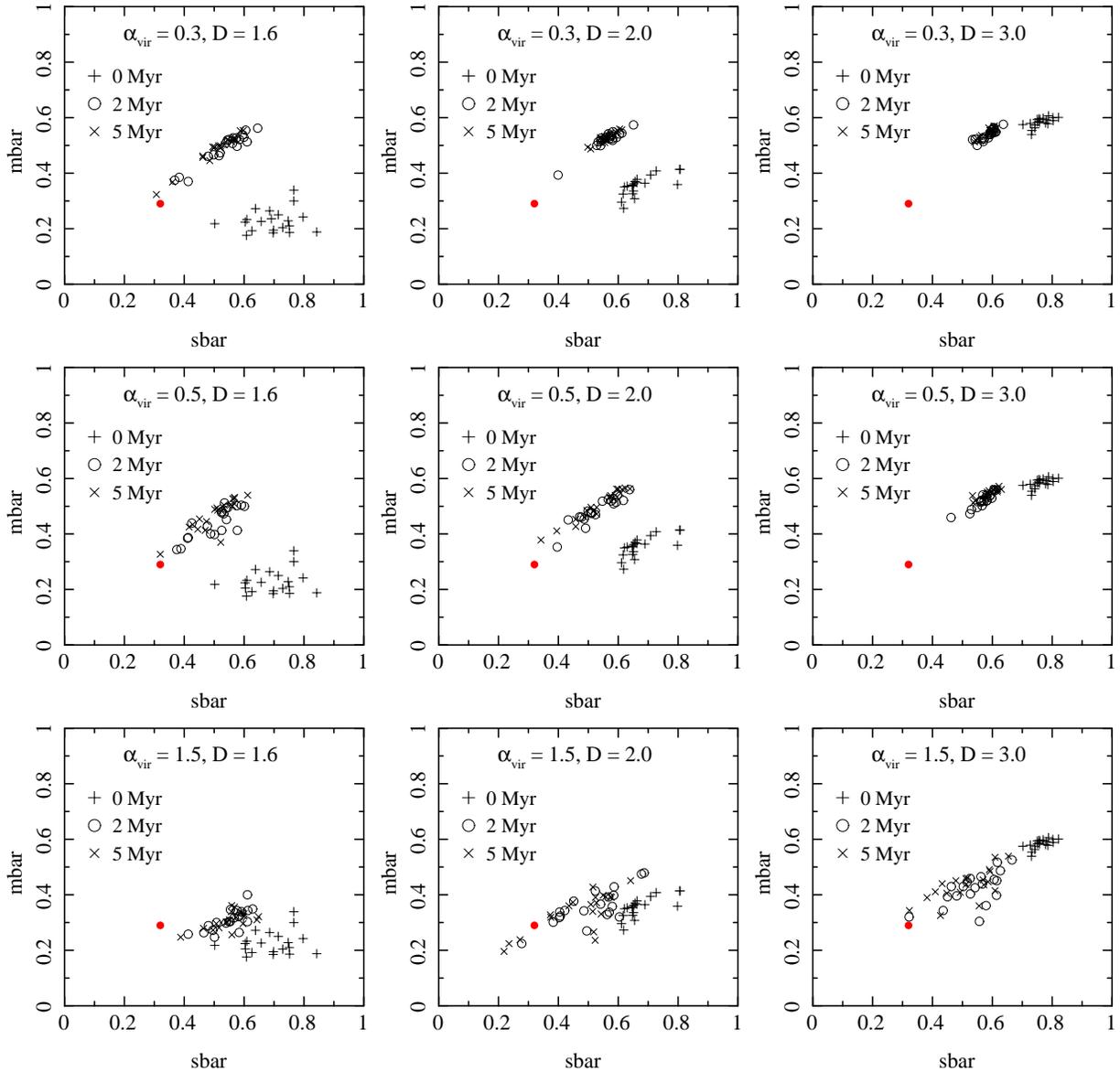}
\caption{Evolution of the $\bar{m}$ (mbar) and $\bar{s}$ (sbar) quantities that constitute the parameter $Q = \bar{m} / \bar{s}$, for all stars within 2 half-mass radii of the centre of each region. For each set of initial conditions we show the values from 20 realisations of the same simulation, and plot values at 0~Myr (before dynamical evolution, plus symbols), at 2~Myr (the estimated age of NGC~6530, open circles), and at 5~Myr (cross symbols). The red dot shows the measured values of $\bar{m} = 0.29$ and $\bar{s} = 0.32$ for NGC~6530.}
\label{mbar_sbar}
\end{center}
\end{figure*}

\subsection{Structural and kinematic diagnostics}

In previous studies \citep[e.g.,][]{park14,wrig14b} we have found that structural diagnostics can also provide insights into the initial conditions of star forming regions, clusters and associations. For example, the $Q$ parameter \citep{cart04} can be used to quantify the amount of spatial structure in a region  by generating a minimum spanning tree (MST) from the 2-dimensional distribution of stars on the plane of the sky \citep{cart04,cart09}. The $Q$ parameter is calculated from the ratio of the normalised mean edge length, $\bar{m}$, and the normalised mean separation between stars, $\bar{s}$, as $Q = \bar{m} / \bar{s}$. Values of $Q < 0.8$ indicate a substructured distribution \citep[such as seen in young star-forming regions or OB associations,][]{cart04,wrig14b}, while values of $Q > 0.8$ are associated with smooth, centrally concentrated distributions such as clusters \citep{park14}. In some cases the values of $\bar{m}$ and $\bar{s}$ can also be useful for understanding the structure of a region.

Figure~\ref{mbar_sbar} shows how the simulations evolve from the different initial conditions in the $\bar{m}$--$\bar{s}$ diagram, using only stars within 2 half-mass radii of the cluster centres. At $t = 0$~Myr (the initial conditions) the distribution in the diagram is only dependent on the initial structure (set by the $D$ parameter). Smoothly distributed groups ($D = 3.0$) start around $(\bar{m}, \bar{s}) = (0.6, 0.8)$ and over time evolve towards lower values of $\bar{m}$ and $\bar{s}$, with the evolution more enhanced for `warm' initial conditions ($\alpha_\mathrm{vir} = 1.5$) compared to the `virial' and `cool' initial conditions ($\alpha_\mathrm{vir} = 0.3$ or 0.5). Substructured groups ($D = 1.6$ or 2.0) start around $(\bar{m}, \bar{s}) = (0.3, 0.7)$, with a larger spread, and evolve towards larger values of $\bar{m}$ and smaller values of $\bar{s}$. This evolution is more pronounced for `cool' and `virial' initial conditions ($\alpha_\mathrm{vir} = 0.3$ or 0.5) than for `warm' initial conditions ($\alpha_\mathrm{vir} = 1.5$), which show very little increases in $\bar{m}$, though $\bar{s}$ does still decrease.

For NGC~6530, using our spatially-unbiased structural sample, we measure values of $\bar{m} = 0.29$, $\bar{s} = 0.32$, and therefore $Q = 0.96$. This value of $Q$ suggests NGC~6530 is centrally concentrated and not substructured, which is consistent with the observed spatial distribution of sources that shows a clear, central cluster and a radial decrease in the density of sources. Figure~\ref{mbar_sbar} shows how these measurements compare to the results of N-body simulations. These structural measurements are consistent with N-body simulations that undergone cool collapse from an initially substructured distributed ($D = 1.6$ and $\alpha_\mathrm{vir} = 0.3$ or 0.5), as well as some of the initially `warm' simulations ($\alpha_\mathrm{vir} = 1.5$), though these are ruled out by the comparison with the velocity dispersion ratio (Section~\ref{s-vdr_simulations}).

Taking the results of the structural and kinematic diagnostics together suggests that NGC~6530 must have undergone a cool collapse (with an initial `cool' velocity dispersion, $\alpha_\mathrm{vir} = 0.3$ or 0.5) from an initially more dispersed and substructured spatial distribution ($D = 1.6$) that likely represents its initial formation state.

\section{Discussion}

We have found strong evidence that the PM velocity dispersion of stars in NGC~6530 increases with stellar mass; the most massive stars moving faster than the low-mass stars. Since this was calculated from PMs and not from radial velocities this cannot be due to inflation of the velocity dispersion from instantaneous velocity measurements for stars in binary systems. We therefore conclude that this signal is real and consider possible explanations for it.

An increase of velocity dispersion with stellar mass is the opposite signature to that expected if energy equipartition were being developed \citep[i.e., if two-body encounters were driving the system towards a thermal velocity dispersion,][]{spit69}. Energy equipartition has long been considered to be the dynamical process behind mass segregation, the apparent over-concentration of the most massive stars in the cluster centre \citep[e.g.,][]{hill98}.

However, recent N-body simulations have shown that the most massive stars in a system are not necessarily moving slower than their low-mass counterparts and can even obtain higher velocities than the low-mass stars \citep{park16b,sper16,webb17}. This is thought to happen as a consequence of the mass segregation that leads to the Spitzer instability \citep{spit69}, wherein a stellar system composed of two populations of stars with significantly different masses is prevented from achieving energy equipartition because the massive stars cannot transfer enough energy to the low-mass stars and instead kinematically decouple from them to form a self-gravitating system within the cluster \citep{alli11}. This sub-system is more centrally concentrated than that of the other stars, and since velocity dispersion decreases with increasing distance from the centre of the cluster, the massive stars achieve a higher velocity dispersion than the low-mass stars \citep{webb17}. This central system relaxes by ejecting one or more of the massive stars from the centre through close encounters \citep{alli11}, which can reduce the observed level of mass segregation.

\citet{sper16} also found that the velocity dispersion of the most massive stars ($> 20$~M$_\odot$) tended to be higher than the velocity dispersion of the less-massive stars. In their simulations, which start from a centrally-concentrated distribution, this velocity dispersion signature occurred after approximately 30~Myrs, whereas \citet{park16} find that, in simulations where the stellar system undergoes some degree of violent relaxation, the most massive stars can obtain higher velocities than the low mass stars within 1--2~Myr (see their Figure~6 panels (a) and (d) and our Figure~\ref{ee_simulations}). This timescale is close to the age of the Lagoon Nebula population and so this picture is consistent with our observations.

These results suggest that the NGC~6530 cluster must have formed from an initially substructured spatial and kinematic distribution that underwent a cool collapse, accelerating the dynamical processes responsible for the Spitzer instability. From the N-body simulations presented by \citet{park16} and \citet{sper16} this provides the only explanation for the elevated velocity dispersion for the most massive stars in the cluster. The lack of significant kinematic substructure \citep{wrig19} can be explained if the cluster has undergone rapid mixing following its collapse. Since such a collapse is more likely to have been asymmetric than perfectly symmetric then this would lead to an asymmetric balance of velocities that, post-collapse and bounce, could explain the asymmetric expansion pattern currently observed across the Lagoon Nebula \citep{wrig19}.

The wider implications for this result are particularly interesting, especially if this is a common formation mechanism for young clusters \citep{bonn03,vazq17}. The question of how star clusters form has significant ramifications for models of high- and low-mass star formation and the origin of the clustering of stars. NGC~6530 contains many massive O-type stars that must have formed in a pre-collapse substructured distribution or possibly have accreted additional mass during the collapse process. Close encounters between stars would be more common during the cool collapse of the cluster, which can lead to the truncating of circumstellar discs and the hardening of existing binaries. \citet{bonn03} estimate that at least a third of all stars, and most massive stars, could undergo such disruptive interactions during the cool collapse of the cluster. Furthermore, while NGC~6530 is an order of magnitude less massive than some of the massive star clusters in our Galaxy, such as Westerlund 1 or NGC~3603, it may provide an indication of how such massive young clusters could form \citep{long14}. Similar studies of such clusters may elucidate this matter further.

\section{Conclusions}

We have studied the mass-dependence of the velocity dispersion of stars in the young cluster NGC~6530 and shown that the PM velocity dispersion increases with stellar mass. We show this by fitting the velocity dispersion in different mass bins, by performing a global velocity dispersion fit with a mass-dependent free parameter, and by measuring a new kinematic diagnostic, the velocity dispersion ratio, $\sigma_{\mathrm{VDR}}$. All of these methods show that the velocity dispersion is higher for the most massive stars in the cluster ($\gtrsim 20$~M$_\odot$) than it is for less-massive stars.

This trend of increasing velocity dispersion with stellar mass is the opposite to that predicted for the development of energy equipartition through two-body encounters. It is however consistent with recent N-body simulations that show such a phenomena can be developed over time due to the Spitzer instability in which the massive stars sink to the centre of a cluster and form a self-gravitating system with a higher velocity dispersion. While this kinematic trend can take several tens of Myr to develop in a normal cluster, if the cluster formed by the cool collapse of an initially substructured distribution it can be achieved in only 1--2~Myr \citep{park16}. We therefore conclude that NGC~6530 formed from much more extended initial conditions and has since collapsed to form the cluster we see now.

\section{Acknowledgments}

NJW acknowledges an STFC Ernest Rutherford Fellowship (grant number ST/M005569/1). RJP acknowledges support from the Royal Society in the form of a Dorothy Hodgkin Fellowship. This work has made use of results from the European Space Agency space mission {\it Gaia}. {\it Gaia} data are being processed by the {\it Gaia} Data Processing and Analysis Consortium (DPAC). Funding for DPAC is provided by national institutions, in particular the institutions participating in the {\it Gaia} Multi-Lateral Agreement (MLA). This paper also makes use of public survey data (programme 177.D-3023, the VST Photometric H$\alpha$ Survey of the Southern Galactic Plane and Bulge) obtained at the European Southern Observatory. This research has made use of NASA's Astrophysics Data System and the Simbad and VizieR databases, operated at CDS, Strasbourg. The authors would like to thank Rob Jeffries for discussions on this work and the anonymous referee for careful reading of this paper and helpful suggestions to improve the work.

\vspace{-0.4cm}

\bibliographystyle{mn2e}
\bibliography{/Users/nwright/Documents/Work/tex_papers/bibliography.bib}
\bsp

\end{document}